# Production of oriented nitrogen-vacancy color centers in synthetic diamond

A. M. Edmonds, U. F. S. D'Haenens-Johansson and M. E. Newton
*Department of Physics, University of Warwick, Coventry, CV4 7AL, UK*

K.-M. C. Fu
*Departments of Physics and Electrical Engineering, University of Washington, Seattle, WA 98195, USA*

C. Santori and R. G. Beausoleil
*Hewlett-Packard Laboratories, 1501 Page Mill Rd., Palo Alto, CA 94304, USA*

D. J. Twitchen and M. L. Markham
*Element Six Ltd., King's Ride Park, Ascot, Berkshire, SL5 8BP, UK*

**Abstract**

The negatively charged nitrogen-vacancy ($NV^-$) center in diamond is an attractive candidate for applications that range from magnetometry to quantum information processing. Here we show that only a fraction of the nitrogen (typically < 0.5 %) incorporated during homoepitaxial diamond growth by Chemical Vapor Deposition (CVD) is in the form of undecorated $NV^-$ centers. Furthermore, studies on CVD diamond grown on $(110)$ oriented substrates show a near 100% preferential orientation of $NV^-$ centers along only the $[111]$ and $[\bar{1}\bar{1}1]$ directions, rather than the four possible orientations. The results indicate that NV centers grow in as units, as the diamond is deposited, rather than by migration and association of their components. The NV unit of the $NVH^-$ is similarly preferentially oriented, but it is not possible to determine whether this defect was formed by H capture at a preferentially aligned NV center or as a complete unit. Reducing the number of NV orientations from 4 orientations to 2 orientations should lead to increased optically-detected magnetic resonance contrast and thus improved magnetic sensitivity in ensemble-based magnetometry.

One of the most intensively studied atom-like solid state systems is the negatively charged nitrogen-vacancy (NV⁻) color center in diamond. The NV⁻ photoluminescence (PL) intensity is strongly modulated depending on whether the system is in the $M_S = \pm 1$ or $M_S = 0$ ground electron spin state, facilitating optically detected magnetic resonance on single defects at room temperature. An essentially spin free lattice can be produced by chemical vapor deposition (CVD) of diamond depleted in $^{13}$C and with a low (<$10^{13}$ cm$^{-3}$) concentration of paramagnetic impurities. In this material electron spins of single NV⁻ centers have remarkably long spin de-phasing times ($T_2$) exceeding a few ms even at room temperature. [1] This enables quantum logic elements to be realized utilizing long-range magnetic dipolar coupling between individually addressable single electron spins associated with separate NV⁻ centers in diamond, [2] and nanoscale spin imaging with a single NV⁻ center. [3] It is not only single NV⁻ centers that are attracting attention. Recently the coupling of a superconducting qubit to a spin ensemble of NV⁻ centers has been reported, highlighting the possibility of diamond spin-ensemble based quantum memory for superconducting qubits. [4]

Both neutral (NV⁰) and negatively charged NV centers are typically present in as-grown CVD diamond, unless considerable efforts are made to exclude nitrogen from the source gases. Currently the key exploitable properties of grown-in NV⁻ centers are typically far better than those centers produced by ion implantation. Low yield from implantation could be due to the limited number of vacancies around the implanted nitrogen [5] whereas the sub-optimal spin and optical properties are most probably related to residual implantation damage defects in the vicinity of the NV⁻ center. Revealing the mechanism for NV center production during CVD diamond growth and demonstrating that the orientation can be controlled is the focus of this paper.

We have two possible mechanisms for the production of NV centers during CVD diamond growth. At typical CVD diamond growth temperatures (e.g. 900 – 1500 K) the substitutional nitrogen donor ($N_S$) is immobile while vacancies are highly mobile and can diffuse through the lattice. [6] Hence, if vacancies are injected during growth then NV centers could be produced by vacancy capture at a previously incorporated $N_S$ (known to be very efficient vacancy trap [6]). Alternatively, NV centers could be grown-in as units, with their production related to processes occurring on the growth surface. Nitrogen is readily incorporated into CVD diamond in the form $N_S$ [7] and there is a lattice distortion associated with $N_S^0$ due to the occupation of a N—C anti-bonding orbital by the *extra* electron (compared to carbon) of nitrogen. [8] The unique N—C bond is about 25-30 % longer than the normal C—C bond length and the donor level is very deep ($E_A$ = 1.7 eV). [9] However, when empty or partially empty states are available in the band gap (e.g. NV⁰) the extra nitrogen electron can be transferred to the available state (e.g. producing NV⁻), the lattice distortion disappears and the positively charged substitutional nitrogen center ($N_S^+$) [10] sits on the tetrahedral lattice site. In addition to the two charge states of $N_S$ and NV defects, the nitrogen-vacancy-hydrogen complex (see supplementary information) is a common defect in CVD diamond, existing as NVH⁻ or NVH⁰. [11-13] Thermal treatment (at temperatures well below the growth temperature and hence unlikely to result in any structural changes) and ultraviolet irradiation can significantly change the optical absorption of nitrogen doped CVD diamond and the concentrations[1] of $N_S^{0/+}$, $NV^{0/-}$ and $NVH^{0/-}$.

---

[1] Square brackets as in [X] are used to indicate the concentration of X.

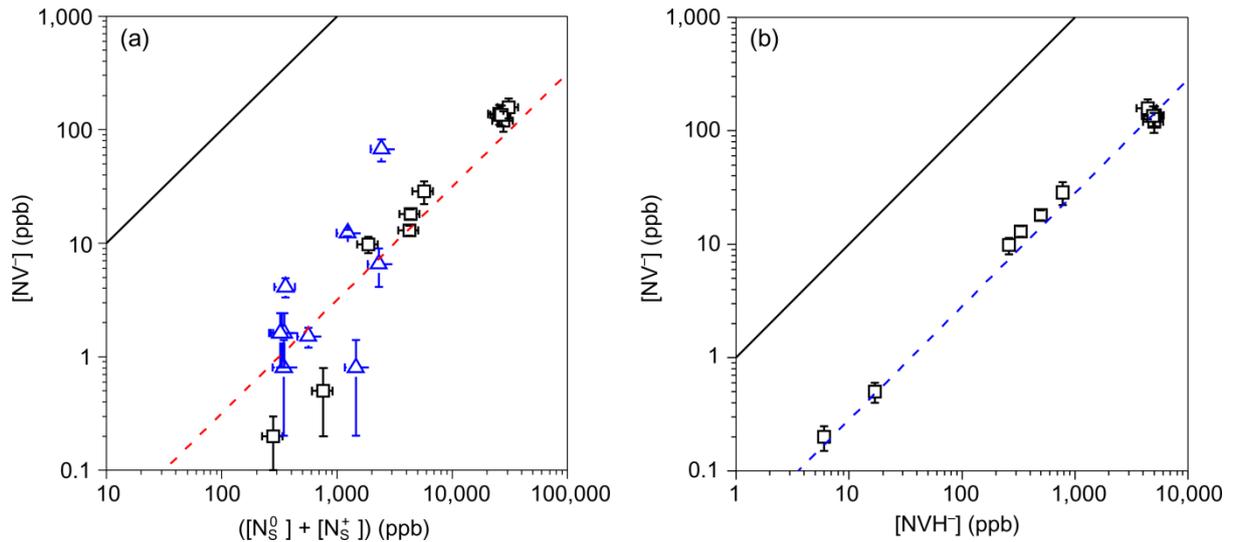

Figure 1: (a) The NV⁻ concentration plotted against the total substitutional nitrogen concentration in diamond samples grown by CVD. The solid line is plotted for the situation where $([N_S^0] + [N_S^+]) = [NV^-]$. (b) The concentration of NV⁻ centers plotted against the concentration of the NVH⁻ centers. The solid line is plotted for the situation where $[NV^-] = [NVH^-]$. In both cases the broken line is a $y = mx$ fit to the data points. In all figures the data points marked with □ were obtained from measurements on samples grown by Element Six Ltd and deliberately doped with nitrogen, and those marked with △ were on samples purchased from Apollo Diamond Inc.

The data in Fig. 1(a) show the metastable (measured after prolonged exposure to laboratory/day light at room temperature) concentration of NV⁻, measured by Electron Paramagnetic Resonance (EPR), plotted against the sum of the neutral ($[N_S^0]$, measured by EPR) and positively charged ($[N_S^+]$, measured by IR absorption) $N_S$ defect concentrations, ($[N_S^+] + [N_S^0]$), in CVD diamond samples from different suppliers, but all grown on approximately {100} oriented substrates. [NV⁻] is typically more than two orders of magnitude less than the concentration of $N_S$. Although there is considerable scatter in the data, the trend is that [NV⁻] increases with total substitutional nitrogen content. In the highest purity intrinsic material studied $[N_S^0] \approx 0.1$ ppb (measured by EPR; detection limit for 10 mm³ samples is ~0.02 ppb). Taking a value of 0.1 ppb for the total $N_S$ defect concentration in the highest purity material with the data in Fig. 1(a) suggests that the NV⁻ concentration is $\leq (0.2\text{-}0.5) \times 10^{-3}$ ppb in such material. This corresponds to on average $\leq$ 10-30 NV⁻ centers in a volume of 10×10×4 μm³, a value which is consistent with NV counting in such material. [14] Figure 1(b) shows [NV⁻] plotted against [NVH⁻] for the samples (measured in the metastable state) grown by Element Six Ltd on (100) oriented substrates, where the concentration of both defects could be measured by EPR. There is a strong correlation between [NV⁻] and [NVH⁻], with the latter being greater than an order of magnitude more abundant. This may indicate that under controlled conditions the ratio of [NV⁻] to [NVH⁻] is fixed over a wide range of nitrogen doping. Data on samples from other suppliers follow this trend, but with much greater scatter. The ratio of concentrations ($[N_S^+] + [N_S^0]$):[NVH⁻]:[NV⁻] is typically 300:30:1 in nitrogen doped CVD diamond samples grown on (100) oriented samples. In the limited number of single growth sector samples studied that were grown on (110) oriented substrates the ratios between the different defect concentrations are similar, but the concentration of $N_S^0$ is approximately three times higher in the samples grown on (110) oriented substrates than on (100) oriented substrates under identical growth conditions. Samlenski et al. [15] showed that nitrogen was more readily incorporated into the {111} growth sectors than {100} sectors, by a factor of between three and four, but noted, as has subsequently been reported by others, [16] that there is a low incorporation probability for the nitrogen dopant.

The $N_S^0$, $NV^0$ and $NV^-$ defects are known to have $C_{3v}$ symmetry and $NVH^-$ motionally averages to an effective $C_{3v}$ symmetry on the EPR time scale. [12] For a diamond sample homoepitaxially grown on an (001) oriented substrate (or any of the {100} family), all the ⟨111⟩ directions are equivalent with respect to the [001] growth direction. However, for CVD diamond grown on an (110) oriented substrate, two ⟨111⟩ directions are perpendicular to the vertical [110] growth direction (i.e. they lie in the growth plane) and two ⟨111⟩ directions make an angle of approximately 35° with respect to the growth direction (i.e. they point out of the growth plane). Figure 2(a) shows a model hydrogen terminated (110) diamond surface (including a growth step) in which a nitrogen atom has been incorporated. This nitrogen is three-fold coordinated with a non-bonding lone pair pointing out of the surface. When the nitrogen is overgrown it is possible that either a $N_S$ or a NV defect is produced depending on whether a vacancy is incorporated in the next layer *above* the nitrogen. In this way only [111] and [$\bar{1}\bar{1}1$] oriented NV centers could be produced.

It is thought that growth on a flat {110} surface is initiated by H abstraction from and $CH_3$ chemisorption on the diamond surface, H abstraction from and $CH_3$ addition to the chemisorbed $CH_3$ molecule, H abstraction from the second $CH_3$ molecule and from an adjacent site on the diamond surface, and bonding between the $C_2H_2$ radical and the diamond. [17] Thus a nucleus for the next layer growth is formed (i.e. an island has formed by bridging two chemisorbed C atoms across an atomic "trough" on a {110} surface), and subsequent monolayer growth requires only chemisorption of one C atom. Therefore, the "trough-flow" growth of {110} layers is likely to be much more rapid than their nucleation. If a N atom is incorporated into this "trough-flow", the site adjacent to the N atom is very attractive for C addition since the chemisorbed atom would form two bonds with the surface, whilst the N atom adopts a three-fold coordination. Hence it seems very unlikely that [$1\bar{1}\bar{1}$] or [$\bar{1}1\bar{1}$] oriented NV centers would be formed for growth on a (110) surface. For "trough-flow" growth on a layer above an incorporated N atom, when the growth front reaches the three-fold coordinated N the probability for addition of a C atom should be somewhat reduced and hence a [111] and [$\bar{1}\bar{1}1$] oriented NV unit introduced for growth on a (110) surface. Theoretical modeling of this process is likely to be challenging, but it would be very interesting to see if this intuitive explanation and the reported experimental data are supported by detailed calculation.

One cannot observe preferential orientation of the $N_S^0$ center, since it is known that this center rapidly re-orientates between the different symmetry related ⟨111⟩ distortions, [18, 19] even at room temperature. Ensemble EPR studies on the incorporation of $NV^-$ defects into a free standing single sector homoepitaxial CVD diamond sample grown on an (110) oriented substrate are reported in Figs. 2(b) and 2(c). White light optical excitation was used to spin polarize $NV^-$, dramatically increasing sensitivity. [20] Fig. 2(b) shows the EPR spectrum recorded from this sample with the magnetic field close (± 1°) to the [$1\bar{1}\bar{1}$] crystallographic direction (i.e. approximately perpendicular to the ⟨110⟩ growth direction), and in Fig. 2(c) the EPR spectrum recorded with the magnetic field was oriented approximately (± 1°) along the [111] crystallographic direction.

Unsurprisingly the $N_S^0$ EPR spectra are identical, independent of which $\langle 111 \rangle$ direction the magnetic field is oriented along, as expected for a defect which is aligned along the four possible orientations with equal probability. However, it is clearly seen that the NV⁻ EPR spectra differ for the two orientations of the magnetic field. When the magnetic field is aligned along $[1\bar{1}\bar{1}]$ (Fig. 2(b)) the outermost EPR lines from NV⁻ centers aligned parallel to the magnetic field (visible in Fig. 2(c)) are absent; there are no detectable NV⁻ centers oriented along the $[1\bar{1}\bar{1}]$ direction. Similarly it can be shown that there are no NV⁻ centers oriented along the $[\bar{1}1\bar{1}]$ direction. When the magnetic field is aligned along $[111]$ (Fig. 2(c)) or $[\bar{1}\bar{1}1]$ the outermost EPR lines resulting from NV⁻ centers aligned parallel to the magnetic field are present. It must be remembered that the spin-polarization of the NV⁻ ground state depends on the orientation (and magnitude) of the magnetic field with respect to the defect axis and is at a maximum when the two are parallel. This is the reason that the inner NV⁻ transitions in Fig. 2(c) are less intense then the outer pair. Analysis of the EPR peak intensities in Fig. 2 reveals that > 97 % of the NV⁻ centers are oriented along either the $[111]$ or $[\bar{1}\bar{1}1]$ directions. If a vacancy migrated to a pre-existing $N_S$ defect all orientations of the NV center would be possible so the data clearly indicate that the NV center is grown-in as a unit. The NVH⁻ centers are similarly preferentially oriented with respect to the growth direction, indicating that NVH⁻ centers are grown-in as complete units or at least the preferentially oriented NV fragments are grown-in and later each capture a mobile hydrogen atom.

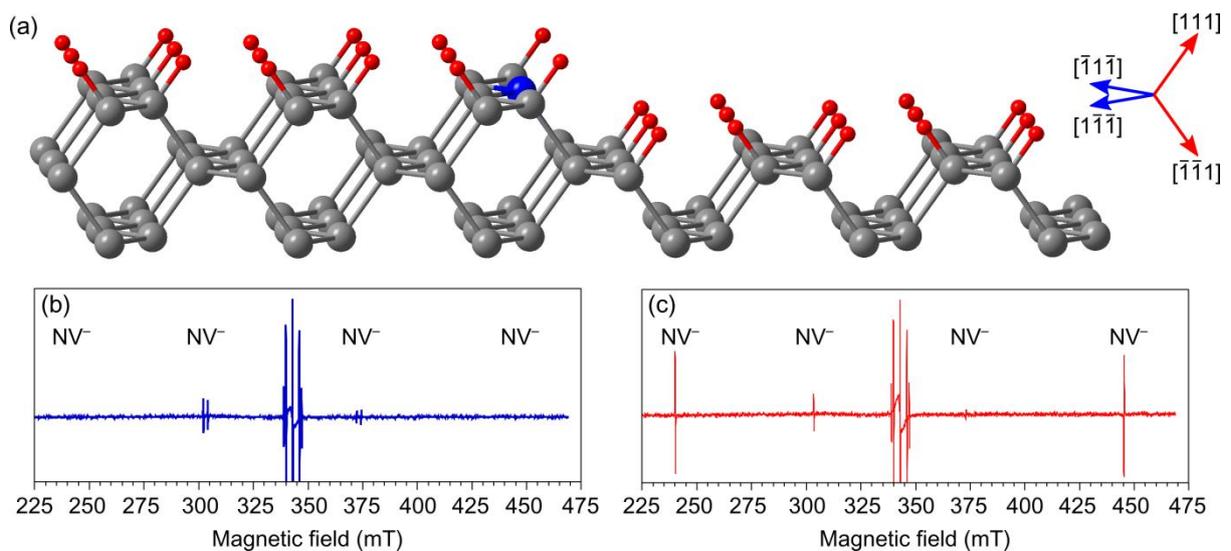

Figure 2: (a) Idealized representation of a hydrogen terminated (110) diamond surface (including a growth step) in which a nitrogen atom (shown without a surface hydrogen) has been incorporated. Inset shows the orientation of the possible $\langle 111 \rangle$ directions. (b) and (c) RT 9.5 GHz CW EPR spectra from (110)-grown CVD diamond sample. The narrow lines close to the center of the spectrum originate from the $N_S^0$ defect and those further from the center from the NV⁻ defect. White light optical excitation from a HgXe arc lamp was used to spin polarize the EPR spectrum from the NV⁻ defect and increase sensitivity. In (b) the EPR spectrum was recorded with the magnetic field oriented close ($\pm 1°$) to the $[1\bar{1}\bar{1}]$ crystallographic direction (i.e. approximately perpendicular to the $[110]$ growth direction) and in (c) the magnetic field was oriented approximately ($\pm 1°$) along the $[111]$ crystallographic direction.

Ensemble EPR measurements showed that in samples containing relatively high levels of nitrogen impurities the NV centers were preferentially grown in as units. In order to confirm that this was also true for samples with much lower concentrations of defects consistent with decoherence times in excess of 0.5 ms, high-purity epitaxial synthetic diamond layers grown on an (110) oriented diamond substrate were characterized using confocal PL imaging. Two 40×40 µm confocal PL images taken 38 µm beneath a diamond surface are shown in Figs. 3(a) and 3( b). This depth was chosen to minimize background fluorescence from the surface in the final images however single NVs can also be detected at the surface within the optical depth of focus (4 µm). The confocal images are composite images obtained from summing 9 separate scans in which laser polarization is encoded into three color channels according to the methodology discussed previously. [21] Because the excitation rate and collection efficiency depend on the relative angle between the laser polarization and the NV axis, centers with different orientations appear as spots with different colors in the composite image. This is clearly seen in Fig. 3(a) for a natural diamond sample with a (111)-polished surface. For the (110)-grown sample shown in Fig. 3(b), single NV$^-$ centers are again clearly resolved, but all of them appear to have approximately the same excitation polarization dependence, indicating preferential orientation. From Fig. 3(b) and similar data we estimate that this sample contains approximately 10 NV$^-$ centers in 400 µm$^3$ which is in accord with the predictions based on the data in Fig. 1, assuming that $[N_s^0] \approx 0.1$ ppb in this material.

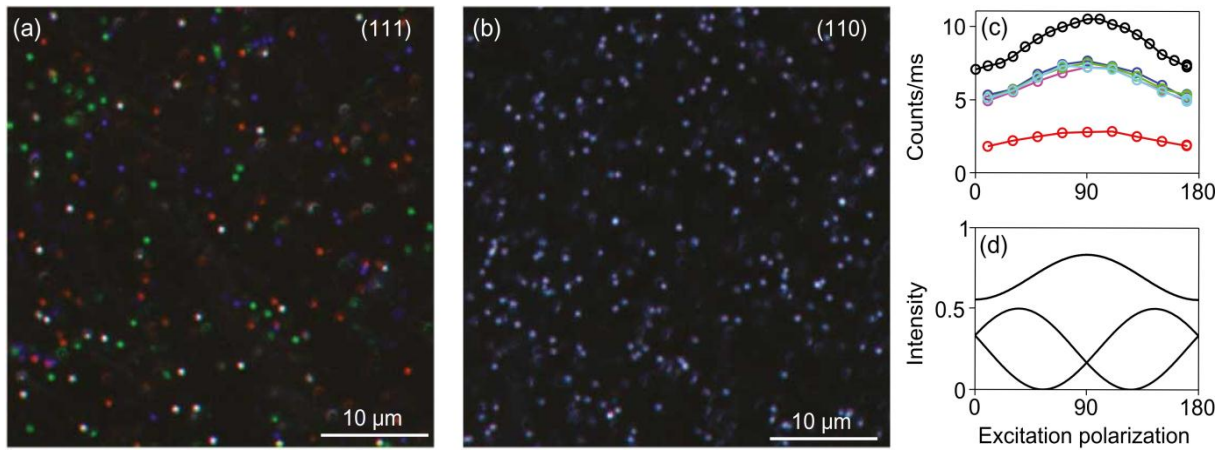

Figure 3: (a) Confocal scanning photoluminescence (PL) image of randomly-oriented single NV centers in a (111)-polished natural diamond sample. This composite image was generated by encoding laser polarization dependence into color as described in Ref. [21], such that the four possible N-V angles are clearly distinguished. (b) A similar image obtained from a diamond homoepitaxial layer grown on a (110) oriented diamond substrate. All of the observed NV centers have similar polarization dependence. (c) Polarization dependence of the PL intensity for 6 NV centers in the (110) sample. (d) Theoretical room-temperature polarization dependence for in-plane (lower curves) and out-of-plane (upper curve) NV centers. The maximum intensity 1 corresponds to a center with a dipole moment aligned to the excitation field and perpendicular to the collection/excitation axis. The model does not include saturation effects.

In Fig. 3(c), the PL intensity is plotted vs. laser polarization angle for six selected NV centers in the (110) sample, with 0° corresponding to an electric field along [001]. Variation in overall brightness may be caused by some NV centers being out of focus, or due to contributions from multiple NV centers. However, the observed contrast ($I_{max}-I_{min})/I_{max}$, in which $I_{max(min)}$ is the maximum (minimum) intensity in the polarization curve, varies within a small range, 0.32-0.36. Figure 3(d) shows the theoretically expected dependence for in-plane (lower two curves) and out-of-plane (upper curve) orientations. The calculation assumes that the excitation power is far below saturation, and that the two excited orbital states are equally populated at room temperature. [22] The theoretical contrasts are 1/3 and 1 for out-of-plane and in-plane NV centers, respectively. The observed polarization dependence is consistent only with that expected for NV centers oriented out-of-plane.

Our analysis of a series of PL images found only a few objects out of approximately 1200 examined with anomalous polarization dependence. Upon further investigation, these spots were found not to be associated with stable NV defects. Thus we estimate that >99.9% of the NV centers in the studied region are preferentially oriented with their trigonal axes pointing out of the (110) growth plane.

Preferential orientation of the NV$^-$ and NVH$^-$ centers has been observed in samples grown on (110) oriented substrates over a wide range of nitrogen doping (> 3 orders of magnitude) and interpreted in terms of the NV unit growing in as the diamond is deposited rather than by migration and association of their components. Although it is not possible to confirm that the mechanism for incorporation of NV$^-$ and NVH$^-$ centers is the same on (001) oriented surfaces the similar ratio of concentrations of $N_S^0$, NVH$^-$ and NV$^-$ defects in (001) and (110) oriented grown CVD diamond samples supports such an assertion. The experimental data unambiguously show that only a small fraction (typically less than 0.5 %) of the nitrogen is incorporated as the undecorated NV unit, with the vast majority as $N_S$. The NV concentration is often so low that if during growth there were even a very small concentration of mobile vacancies then the NV's produced by vacancy capture at the much more abundant $N_S$ defect would dominate and no preferential orientation of NV's would be observed. Hence we propose, at least for the samples studied here, there are very few mobile vacancies present during growth. In nitrogen doped material the combined concentration of NVH$^-$ and NV$^-$ is typically very much less than the concentration of $N_S^+$, indicating the presence of as yet unidentified traps. [13]

Reducing the number of NV orientations from 4 orientations to 2 orientations is a significant advancement and should lead to increased optically-detected magnetic resonance contrast and thus increased magnetic sensitivity in ensemble-based magnetometry. For quantum information applications control of the orientation of the NV optical dipole will be important for NV-cavity coupling. For magnetometry applications the highest sensitivity magnetometers will be obtained in ensemble systems in which all NVs are similarly oriented. Preferential orientations of NV centers created by implantation and annealing has not been demonstrated but near 100 % out of plane orientation is possible for homoepitaxial growth on an (110) oriented diamond substrate. In the samples studied we are able to observe NVs at the surface within the optical depth of focus (approximately 4 μm). Future directions for this work include creating preferentially oriented NVs very close to the diamond surface (10-100 nm) and suppressing the excitation of one of the two orientations through a combination of the optical excitation polarization and diamond orientation.

**Methods**
**Electron Paramagnetic Resonance**: Quantitative ensemble room temperature continuous wave Electron Paramagnetic Resonance (EPR) measurements were carried out using commercially available Bruker EMX and ELEXSYS EPR spectrometers operating at approximately 9.5 GHz (X-band). The systems were set up so that it was possible to rotate the sample in two perpendicular planes to achieve a desired orientation with respect to the magnetic field. Microwave power saturation, which occurs when the spin lattice relaxation rate is not sufficiently high to maintain the equilibrium spin population distribution while stimulated transitions are excited by microwaves, needs to be considered if EPR is to be used in a quantitative manner. The spectrometers were run using magnetic field modulation such that the spectral features approximate the first derivative of the EPR line shape. The EPR intensity was determined by fitting the experimental spectrum to a simulated spectrum, deconvolving overlapping spectra from different defects, and integrating the latter twice using a computer program developed in-house. A Tsallis function was used to produce the simulated spectra since EPR line shapes are usually not well reproduced with Lorentzian or Gaussian functions. Furthermore, the algorithm utilizes the pseudo-modulation technique to account for the distortion of the EPR line shape due to field modulation. Defect concentrations were calculated by comparing the EPR signal intensities to that of a reference sample of known concentration. The reference sample used in this study is a small, single growth sector, HPHT synthetic type Ib diamond containing 270 ± 20 ppm atoms of $N_S^0$ (1 ppm = 1 part per million carbon atoms = spin density of $1.76 \times 10^{17}$ cm$^{-3}$). With signal averaging the room temperature EPR detection limit in a diamond sample of volume ≈ 10 mm$^3$ for [NVH$^-$] is approximately 0.2 ppb, whereas that for [$N_S^0$], utilizing rapid passage techniques, is approximately 0.02 ppb. In the dark the ensemble EPR detection limit for NV$^-$ is similar to that for NVH$^-$ but with optically excited spin polarization this can be reduced by two orders of magnitude. [20] However, since a number of factors can influence the spin polarization efficiency quantitative analysis is more challenging. White-light from a HgXe arc lamp directed onto the sample using a liquid light guide (transmission 315-700 nm) was used to spin polarize the negatively charged nitrogen-vacancy centers and improve the EPR detection sensitivity for measurements of the preferential orientation of these color centers.

**Infrared absorption (IR) spectroscopy**: A Perkin Elmer Spectrum GX Fourier Transform IR spectrometer equipped with a beam condenser was used to monitor the concentration of $N_S^+$. At low levels, [$N_S^+$] was estimated from the strength of the infrared absorption at 1332 cm$^{-1}$, rather than using the entire $N_S^+$ infrared absorption band. [10] This process assumes that only $N_S^+$ contributes to the infrared absorption at 1332 cm$^{-1}$ and hence gives an upper limit. However, with this caveat in mind, it does allow an estimation of [$N_S^+$] down to ≈ 200 ppb.

**Confocal Photoluminescence**: Single center room temperature photoluminescence measurements were carried out using a confocal setup in which an excitation laser (532 nm) was focused to a submicron spot inside the sample for non-resonant excitation of the NV centers. This setup also allowed for the control of the laser polarization incident on the sample.

**Acknowledgements**

The staff of the DTC Research Center and Mr Chris Kelly is thanked for their help with sample preparation and characterization. U.F.S.D.J. would like to thank the DTC Research Center for an Industrial Bursary. We acknowledge funding from the FP6 project EQUIND, Defense Advanced Research Projects Agency under Award No. HR0011-09-1-0006 and the Regents of the University of California, the EPSRC and the Science City Research Alliance (SCRA) Advanced Materials projects (AM1 and AM2), which are supported by Advantage West Midlands and in part funded by the European Regional Development Fund.

**Supplementary Information**

# Production of oriented nitrogen-vacancy color centers in synthetic diamond

A. M. Edmonds, U. F. S. D'Haenens-Johansson and M. E. Newton
*Department of Physics, University of Warwick, Coventry, CV4 7AL, UK*

K.-M. C. Fu
*Departments of Physics and Electrical Engineering, University of Washington, Seattle, WA 98195, USA*

C. Santori and R. G. Beausoleil
*Hewlett-Packard Laboratories, 1501 Page Mill Rd., Palo Alto, CA 94304, USA*

D. J. Twitchen and M. L. Markham
*Element Six Ltd., King's Ride Park, Ascot, Berkshire, SL5 8BP, UK*

**Nitrogen-vacancy-hydrogen complex**

The negatively charged nitrogen-vacancy-hydrogen complex $NVH^-$, is also a common defect in nitrogen doped CVD diamond. [1] In this defect the nitrogen is bonded to three carbon atoms with the lone electron pair on the nitrogen (as in the $NV^-$ defect) directed into the vacancy. The hydrogen atom is bonded to one of the carbon atoms surrounding the vacancy, with the unpaired electron localized in an anti-bonding orbital formed between the two carbon dangling orbitals (the bonding orbital being full). [2] The static defect has $C_{1h}$ symmetry, but the hydrogen atom tunnels between the three equivalent $C_{1h}$ configurations (C–H bonds) at a rate that is comparable with the reciprocal Electron Paramagnetic Resonance (EPR) timescale, [2] leading to a motionally averaged EPR spectrum. $NVH^0$ is an electron trap, and recent studies [3] have shown that thermal treatment (at temperatures well below the growth temperature and unlikely to result in any structural change) and ultraviolet irradiation can significantly change the optical absorption of nitrogen doped CVD diamond, and the concentrations of $N_S^0$, $N_S^+$, $NVH^0$ and $NVH^-$. $NVH^0$ is diamagnetic and cannot be detected by EPR but the infrared absorption line at 3123 $cm^{-1}$ has been attributed to a C–H vibrational mode at this defect. [3]

**Preferential Orientation of Nitrogen-vacancy-hydrogen complex**

Figure 1 shows high resolution room temperature EPR spectra, from the same free-standing homoepitaxial CVD diamond sample grown on an $(110)$ oriented substrate discussed in the paper, focusing on the region containing the central transition of the $N_s^0$ defect and the NVH$^-$ spectra. In Fig. 1(a) the EPR spectrum was recorded with the magnetic field oriented along the $[1\bar{1}\bar{1}]$ crystallographic direction and in Fig. 1(c) the magnetic field was oriented along the $[111]$ crystallographic direction. The resulting EPR spectra depend markedly upon whether the magnetic field is oriented along a $\langle 111 \rangle$ direction in or out of growth plane, indicating preferential orientation of the NVH$^-$ centers. The simulations shown in Fig. 1(b) and 1(d) are generated assuming that the NV unit of each of the NVH$^-$ centers are oriented along either the $[111]$ or $[\bar{1}\bar{1}1]$ crystallographic directions (equal probability along either direction) and the $N_s^0$ centers (giving rise to the large central line and $^{13}$C hyperfine satellites) are randomly oriented over all possible $\langle 111 \rangle$ directions. It is assumed that no other defect contributes to the EPR spectra. It is clear that the experimental data is well reproduced by the assumption that the NV unit of each of the NVH$^-$ centers are all preferentially oriented with respect to the growth direction, indicating that NVH$^-$ centers are grown in as complete units, or at least the preferentially oriented NV fragment is grown-in and later captures a mobile hydrogen atom. In fact, the modest signal to noise ratio in Fig. 1 only allows the determination of a lower limit for the preferential orientation, which is > 80 %.

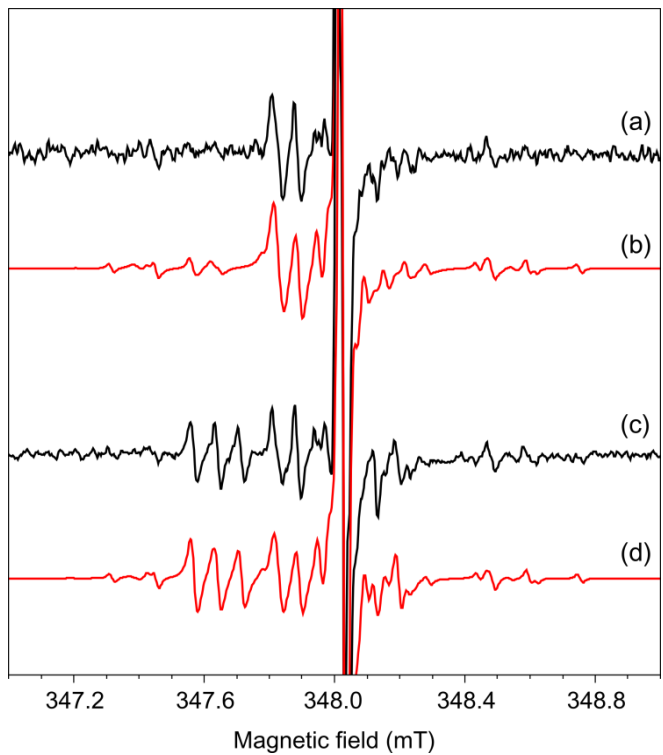

Supplementary Information Figure 1: (a) and (c) Room temperature EPR spectra from a free standing homoepitaxial CVD synthetic diamond sample grown on an (110) oriented substrate (which has been removed) recorded at a microwave frequency of ≈ 9.5 GHz. In (a) the EPR spectrum was recorded with the magnetic field oriented along the $[1\bar{1}\bar{1}]$ crystallographic direction (i.e. perpendicular to the $[110]$ growth direction) and in (c) the magnetic field was oriented along the $[111]$ crystallographic direction. The simulations shown in (b) and (c) are generated assuming that all the NVH⁻ centers are oriented along either the $[111]$ or $[\bar{1}\bar{1}1]$ crystallographic directions (equal probability along either direction) and the $N_s^0$ centers are randomly oriented over all possible $[111]$ directions. In (b) and (d) the magnetic field is parallel to the $[1\bar{1}\bar{1}]$ and $[111]$ crystallographic directions, respectively.